\documentstyle[preprint,aps,version2]{revtex}
\begin{document}

\begin{titlepage} 
\begin{title}
Final-state Interaction Effects on  Inclusive 
Two-particle  Production in 
Electron-positron Annihilation\thanks{This 
work was  supported in part by  the National Science Foundation of China.}
\end{title}

\author{Jian-Jun Yang$^{1,2}$ and  Wei Lu$^{\protect 2,3}$ }

\begin{instit}
1. Department of Physics, Nanjing Normal University, 
Nanjing 210024,  P.R. China

2. Institute of High Energy Physics, 
 P.O. Box 918(4), Beijing 100039, P.R. China

3.  Centre de Physique Th\'eorique,  Ecole Polytechnique, 
91128 Palaiseau Cedex,  France

\end{instit}

\begin{abstract}

The final-state interaction effects on  the 
inclusive  two-particle  production 
in electron-positron annihilation are  investigated
within the context of the one-photon annihilation approximation. 
Such  effects are  characterized by one structure function
in the decomposition of the hadronic 
 tensor.  On the basis of the positivity, 
we derived an inequality to bound  this structure function. 
The price to access it experimentally  is to polarize longitudinally 
one of the initial-state beam, to say, the electron beam, 
and measure  the corresponding single spin asymmetry.
By combining the Callan-Gross relation  with 
our positivity analysis, we obtain an upper bound  for the 
single spin asymmetry  considered. 

\end {abstract}

\end{titlepage} 

 The electron-positron annihilation has  been and will 
continue to be a Holy Land  in the study of the 
hadron production dynamics. Usually, people assumes 
a physical picture for the electron-positron annihilation
into hadrons like this: The electron and positron annihilates into 
a quasi-free quark-antiquark pair via a virtual vector boson,
and the  quark-antiquark pair hadronizes into the particles 
that reach the detector. At high energies,  the final-state particles 
form two jets, roughly collinear with  the parent 
quark-antiquark pair. [Three or more  jets can also be formed, 
but such events are rare in comparison with two-jet 
events.]  Depending  upon how many   and what kind of particles 
are measured, the inclusive particle production in 
electron-positron annihilation supplies us with an ideal 
place in which to measure various quark fragmentation functions. 
Among others, one can measure a  pair of   particles ,  
one in each jet. For this category of reactions, we have the following 
channels  in  mind: $e^+ e^-\to \pi^+  \pi^-  X$, 
$e^+ e^-\to \pi^0  \pi^0 X$,  $e^+ e^-\to p  \bar p  X$,
 $e^+ e^-\to \Lambda \bar \Lambda X$, 
 $e^+ e^-\to B  \bar B  X$, 
 $e^+ e^-\to D \bar D     X$,  and so on. 

 The study of inclusive two-particle production
in electron-positron annihilation  has a   history
of more than 20 years. 
In  the early 1980s, Collins and Soper \cite{jcc} discussed  the 
``almost back-to-back'' two-pion production in 
electron-positron annihilation, indicating  that
it can be employed to measure the transverse momentum 
dependence of the quark  fragmentation function. 
Later,  Fong and Webber \cite{Webber} systematically predicted  the 
two- as well as one-particle distributions in 
electron-positron annihilation,  and pointed out
the importance of measuring two-particle correlations. 
Artru \cite{artru},  and Chen, Goldstein, Jaffe and Ji 
\cite{ji} observed that $e^+ e^-\to \Lambda \bar \Lambda X$ 
can be used to extract  the quark transversity distribution 
function, with the $\Lambda$  and $\bar \Lambda$ polarization 
measured.  Recently, Boer, Jacob and Mulders  \cite{mulders} 
developed a systematic quantum chromodynamics (QCD) factorization  of the 
process of type $e^+ e^-\to h  \bar h  X$, up to order 
$1/Q$, in terms of a full set of quark fragmentation functions.
 Hopefully,  the B-factory under construction 
at KEK can be used  to  study  the above processes 
as well as $e^+ e^-\to B  \bar B  X$.  At the planned 
Tau-Charm factory \cite{tau},  one  will have  much  chance to 
access  $e^+ e^-\to D \bar D     X$.  The advantage in 
choosing two detected particles to be a particle-anti-particle 
pair lies in that the fragmentation function for the 
chosen particle can be related to that for its anti-particle
by the charge conjugation. 
 
 In the naive quark-parton model,  the original 
quark and antiquark  hadronizes independently after 
they are generated, so there would be no correlation between 
 two  hadrons  that belong to two different jets. 
In this work, we look into this matter in the 
quantum field setting.  As a result, we  find that 
due to the final-state interactions, there is a 
structure function describing the correlation between 
two detected particle, though they fall into separate
jets. Such a fact can be easily understood 
within the context of QCD: 
 Two jets originated from the quark 
and antiquark  need to neutralize their colors
in the end of hadronization, so they  cannot fragments independently 
as assumed in the  naive quark-parton model.

 To be specific, we consider 
$$ e^-(k_1) +e^+(k_2) \to h  (p_1)
+ \bar h (p_2)+X,$$
where  two  inclusively detected  hadrons, $h$ 
and $\bar h$,   are in two opposite jets. 
We  work in the c.m. frame, with $\hat z$-axis in the 
traveling direction of $h$  and $\hat y$-axis 
along the normal of its production plane  determined 
by ${\bf k}_1 \times {\bf p}_1$.  
We work with the one-photon annihilation approximation, 
which is applicable up to the intermediate energies. As usual, 
we define 
\begin{equation}
Q^2\equiv -q^2, \quad  z_1\equiv \frac{2 p_1\cdot q}{q^2}, \quad 
 z_2\equiv \frac{2 p_2\cdot q}{q^2}, 
\end{equation}
where $q$ is the momentum 
of the virtual  time-like photon. 
It is straightforward to show that 
the cross section can be expressed as the following 
differential form
\begin{equation}
\frac{d\sigma}{ dz_1 d\cos\theta_1 dz_2 d|{\bf p}_{2\perp}|^2 d\omega } 
=\frac{\alpha^2}{4 \pi^2 Q^3 }\frac{|{\bf p}_1| }{|{\bf p}_{2 \parallel}| }
L_{\mu\nu}(k_1, k_2) W^{\mu\nu}(q, p_1, p_2), 
\label{cross}
\end{equation}
where   $\theta_1$ is the  outgoing angle  of $h$ 
with respect to the electron beam, ${\bf p}_{2 parallel}$ 
and ${\bf p}_{2\perp}$ are  respectively 
 the parallel and perpendicular 
components of ${\bf p}_2$ with respect to ${\bf p}_1$, 
$\omega $  is  the angle spanned by the 
${\bf k}_1\times {\bf p}_1$  and ${\bf p}_1\times {\bf p}_2$ 
planes.  $|{\bf p}_{2 \parallel}|$ is related to  $|{\bf p}_{2 \perp}|$ by 
\begin{equation}
|{\bf p}_{2 \parallel}|= \frac{z_2Q}{2}
\Big[1-\Big(\frac{2}{z_2Q}\Big)^2
\big(
|{\bf p}_{2 \perp}|^2 +M^2 \big)\Big]^{1/2},
\end{equation}
where $M$ is the mass of the hadron.    
$L_{\mu\nu}(k_1, k_2)$ is the  well known  leptonic tensor. 
And all the information 
on strong interaction is incorporated by the hadronic  tensor: 
\begin{equation} 
W^{\mu\nu}(q,p_1,p_2)=\frac{1}{4\pi}\sum_X 
\int d^4\xi 
\exp(iq\cdot \xi) \langle 0| J^\mu (0)
| h(p_1)\bar h(p_2)X\rangle \langle  h(p_1)\bar h(p_2)
X|  J^\nu (\xi)|
0\rangle. 
\end{equation}

It is a common practice to parameterize various hadronic tensors 
in terms of Lorentz invariant 
 structure functions.  To our best knowledge, 
the  general  Lorentz  decomposition  of $W_{\mu\nu}(q,p_1,p_2)$  has not 
yet been reported in the literature.
Taking into the constraints of gauge invariance, Hermiticity, 
parity conservation,  we can  decompose our 
hadronic tensor  as follows:
\begin{eqnarray}
W^{\mu\nu}(q,p_1,p_{2})&
=&\frac{1}{p_1\cdot q} (-g^{\mu\nu}+\frac{q^\mu q^\nu}{q^2})W_1
+\frac{1}{q^2(p_1\cdot q)} 
(p_1^\mu - \frac{p_1\cdot q}{q^2}q^\mu) 
(p_1^\nu - \frac{p_1\cdot q}{q^2}q^\nu)W_2
\nonumber  \\
& & + \frac{1}{q^2(p_1\cdot q)} \left[
(p_1^\mu - \frac{p_1\cdot q}{q^2}q^\mu) p^\nu_{2\perp}
+p^\mu_{2\perp}(p_1^\nu - \frac{p_1\cdot q}{q^2}q^\nu)\right]W_3
+
\frac{p^\mu_{2\perp} p^\nu_{2\perp}}{q^2(p_1\cdot q)} 
W_4 
\nonumber  \\
& & +\frac{i}{q^2(p_1\cdot q)}  \left[
(p_1^\mu - \frac{p_1\cdot q}{q^2}q^\mu) p^\nu_{2\perp}
-p^\mu_{2\perp}(p_1^\nu - \frac{p_1\cdot q}{q^2}q^\nu)\right]
\hat W\label{dec}, 
\end{eqnarray}
where  we  introduced an auxiliary four-momentum 
 $p^\mu_{2\perp}=(0, {\bf p}_{2\perp}, 0). $
Obviously,  $p^\mu_{2\perp}$ satisfies $p^\mu_{2\perp}\cdot q=0$. 
 $W_1$,  $W_2$, $W_3$, $W_4$ and $\hat W$  are 
five independent structure functions. 
The Lorentz tensor associated with $\hat W$ 
is odd under the  naive time-reversal 
transformation. It is this term that incorporates 
the final-state interaction effects on the inclusive two-particle 
production in electron-positron annihilation. 

 Now we set about deriving some constraints among our 
structure functions  on the  basis of  the 
Hermiticity  of electro-magnetic currents. 
The starting point is  that, 
  for an arbitrary  Lorentz vector $a^\mu$, there is always 
\begin{equation} 
W_{\mu\nu}(q,p_1,p_2) a^{\ast \mu} a^\nu \geq 0. \label{poi}
\end{equation} 
Such an inequality can be  termed the positivity of the
hadronic  tensor.

For a matrix to be semi-definite positive, it 
should satisfy the following sufficient and necessary conditions: 
All  of its sub-matrices have  semi-positively finite determinants 
\cite{soffer}. To  investigate the positivity  restrictions on 
the structure functions in our hadronic  tenor, 
we write out explicitly the elements of $W^{\mu\nu}(q,p_1,p_2)$: 
\begin{eqnarray} 
W^{00}&=&W^{01}=W^{02}=W^{03}=W^{10}=W^{20}=W^{30}=0,\\
W^{11}&=&\frac{2}{Q^4z_1} 
(Q^2W_1+ |{\bf p}_{2\perp} |^2 \cos^2\omega  W_4)\\ 
W^{22}&=&\frac{2}{Q^4z_1} 
(Q^2W_1+ |{\bf p}_{2\perp} |^2 \sin^2\omega  W_4) \\
W^{33}&=&\frac{2}{Q^4z_1}
(Q^2W_1+ |{\bf p}_1 |^2  W_2 ) \\ 
W^{12}&=&W^{\ast 21}=
\frac{2}{Q^4z_1} 
|{\bf p}_{2\perp} |^2 \sin \omega  \cos\omega  W_4\\ 
W^{13}&=&W^{31\ast }=
\frac{2}{Q^4z_1} 
|{\bf p}_{2\perp} | |{\bf p}_1 | \cos\omega  (W_3-i \hat W)\\
W^{23}&=&W^{32\ast }=
\frac{2}{Q^4 z_1} 
|{\bf p}_{2\perp} | |{\bf p}_1 | \sin\omega  (W_3-i \hat W).
\end{eqnarray} 
The fact that $W^{\mu\nu}(q,p_1,p_2)$ is at rank  three is 
simply due to the restriction of the current conservation condition.
 Therefore,  we have   the following  nontrivial 
positivity inequalities: 
\begin{equation} 
\left|
\begin{array}{ccc}
W^{11}& W^{12} &W^{13}\\ 
W^{21} &W^{22} & W^{23}\\
W^{31} &W^{32} & W^{33} 
\end{array}
\right|\geq 0,
\end{equation} 

\begin{equation} 
\left|
\begin{array}{cc}
W^{11} & W^{12}\\ 
W^{21} & W^{22} 
\end{array}
\right|\geq 0, 
~\left|
\begin{array}{cc}
W^{22} & W^{23}\\ 
W^{32} & W^{33}\label{x3}
\end{array}
\right|\geq 0,
\end{equation} 
\begin{equation} 
W^{11}\geq 0, 
~W^{22}\geq 0, 
~W^{33}\geq 0.
\end{equation} 
In our parameterization, the above six inequalities yield 
the following inequalities   among the 
structure functions: 
\begin{equation}
W_1 \left[
(Q^2W_1 +|{\bf p}_{2\perp} |^2 W_4)(Q^2 W_1+|{\bf p}_1 |^2W_2)- 
|{\bf p}_{2\perp} |^2 |{\bf p}_1|^2(W_3^2+ {\hat W}^2)\right]  \geq 0,
\label{l0}
\end{equation}
\begin{equation}
W_1 ( 
Q^2W_1 +|{\bf p}_{2\perp} |^2 W_4 ) \geq 0,
\label{l1}
\end{equation}
\begin{equation}
(Q^2W_1 +|{\bf p}_{2\perp} |^2 \sin^2\omega  W_4)
(Q^2W_1+ |{\bf p}_1 |^2W_2)- |{\bf p}_{2\perp} |^2 |{\bf p}_1 |^2
\sin^2 \omega (W_3^2+ {\hat W}^2) \geq 0,
\label{l2}
\end{equation}
\begin{eqnarray} 
Q^2W_1 +  |{\bf p}_{2\perp}|^2 \cos^2\omega  W_4 &\geq & 0, \label{a}\\
Q^2W_1 +  |{\bf p}_{2\perp}|^2 \sin^2\omega  W_4 &\geq & 0, \label{b}\\
Q^2W_1 +  |{\bf p}_1|^2  W_2 &\geq & 0, \label{c}
\end{eqnarray}
Here some notes are in order. 

1) By letting $\sin\omega =0$ in (\ref{a}) or $\cos\omega =0$
in (\ref{b}),  one can  fix the sign of $W_1$:
\begin{equation}
W_1\geq 0. \label{sign}
\end{equation}
Therefore, (\ref{l0}) and (\ref{l1}) can be safely divided by 
$W_1$ without changing the direction of the inequality signs. 
On the other hand, if one assumes $\sin\omega =1$ 
in (\ref{a}) or $\cos\omega =1$ in (\ref{b}),  these two 
inequalities reduce to (\ref{l1}). 

2) Adding (\ref{a}) to (\ref{b}), one will gain a new inequality: 
\begin{eqnarray} 
Q^2W_1 + \frac{1}{2} 
 |{\bf p}_{2\perp}|^2 W_4 &\geq & 0.  \label{99}
\end{eqnarray}

3). By setting $\sin\omega $ to be 1 or 0 in (\ref{l2}),
one will reproduce  (\ref{l0}) and (\ref{l1}), respectively. 

 Some of the above inequalities have simple physical meanings. 
To see this, let us recall that a  generic Lorentz vector can  be 
expanded using a complete set of  four 
$independent$  vectors.   Considering the current conservation 
condition, we choose  one  of them to be  proportional to $q$, 
and the  rest three to be the three polarization vectors of the
virtual photon: 
\begin{eqnarray} 
e^\mu_1&=& -\frac{1}{\sqrt 2} (0, 1, +i, 0),\\ 
e^\mu_2 &= & +\frac{1}{\sqrt 2} (0, 1, -i, 0),\\ 
e^\mu_3&=&  (0, 0, 0, 1). 
\end{eqnarray}
Notice that these three polarization vectors are 
orthonormal, namely, 
\begin{equation} 
e^\ast_i \cdot e_j =- \delta_{ij},{\rm ~with~} i,j=1,2,3
\end{equation}  
In  addition, they satisfy the Lorentz condition 
\begin{equation} 
e_i\cdot q=0,  ~~i=1, 2, 3. 
\end{equation} 
Obviously,  by letting $a=q$ in (\ref{poi})  one has  
 an identity $0\equiv 0$ only, which  reflects the 
current conservation of the electro-magnetic interaction. 

If one takes $a$ to be the transverse photon polarization 
vector,  there will be 
\begin{equation} 
e^{\ast\mu}_1 e^\nu_1 W_{\mu\nu}
=
e^{\ast\mu}_2 e^\nu_2 W_{\mu\nu}
\propto
Q^2 W_1+ \frac{1}{2}|{\bf p}_{2\perp}|^2 W_4.
\end{equation}
On the other hand, $e^{\ast\mu}_1 e^\nu_1 W_{\mu\nu}$ 
and $e^{\ast\mu}_2 e^\nu_2 W_{\mu\nu}$ are proportional to the 
cross sections for the virtual  photon with a
transverse polarization,
 so (\ref{99}) simply indicates  that  these cross 
sections are  semi-positive. 

If taking $a$  to be the longitudinal photon  polarization  vector, 
one  will have 
\begin{equation} 
e^{\ast\mu }_3 e^\nu_3 W_{\mu\nu}
\propto Q^2 W_1 +| {\bf p}_1 |^2 W_2, 
\end{equation} 
so (\ref{c})  implies that  the  cross section is semi-positive 
for the virtual  photon with a longitudinal 
polarization. 

Similarly, if  setting the  polarization of the virtual photon 
to be $(e_1+e_2)/\sqrt{2}$ or $i(e_1-e_2)/\sqrt{2}$, one will
attain (\ref{a}) and (\ref{b}), which means  that the 
cross sections for the fragmentation of circularly  polarized 
photon  are also semi-positive. 

  However, inequality (\ref{l2}), which is of most relevance
to the study of the  final-state interactions,  
has no  simple interpretations. 
Physically, it  reflects the interference effects among the 
fragmentation processes of  the time-like photons with 
different polarization states.

As can be seen from Eq. (\ref{dec}), the 
contributions of the final-state interactions to the 
hadronic  tensor are  imaginary and antisymmetric.  
To access   $\hat  W$,  one needs to polarize 
one of the initial beams. 
We consider the case in which the electron beam is longitudinally 
polarized. [The case of the transversely polarized electron beam is of 
no practical meaning  because the corresponding 
 contribution of $\hat W$ 
to  the cross section  will be 
 proportional to $m_e/Q$.]   Correspondingly, 
the spin state of the incident electron can be characterized by 
its helicity $\lambda$.  As a result, one can deduce the following 
cross section formula 
\begin{eqnarray}
\frac{d\sigma(\lambda)}
{dz_1 d\cos\theta_1 dz_2 d|{\bf p}_{2\perp}|^2 d\omega } 
&=&\frac{\alpha^2}{4 \pi^2 z_1 Q^5 }
\frac{|{\bf p}_1| }{|{\bf p}_{2 \parallel}| }
\Big[ 2Q^2 W_1+ |{\bf p}_1|^2 \sin^2\theta_1 W_2
\nonumber \\
&+ & |{\bf p}_{2\perp}| |{\bf p}_1| \sin 2\theta_1 \cos\omega  
W_3 +|{\bf p}_{2\perp}|^2(1-  \sin^2\theta_1 \cos^2\omega  )W_4 
\nonumber \\
&+ & 4 \lambda |{\bf p}_{2\perp}| |{\bf p}_1| \sin\theta_1 \sin\omega  
\hat W \Big].
\label{cs}
\end{eqnarray}

In principle, one can access $\hat W$ by measuring 
the single longitudinal spin asymmetry  defined as 
\begin{equation} 
A_L\equiv \frac{
\displaystyle
\frac{d\sigma(\lambda= +\frac{1}{2})}
{ dz_1 d\cos\theta_1 dz_2 d|{\bf p}_{2\perp}|^2 d\omega } 
-
\displaystyle
\frac{d\sigma(\lambda= -\frac{1}{2})}
{ dz_1 d\cos\theta_1 dz_2 d|{\bf p}_{2\perp}|^2 d\omega } 
}
{
\displaystyle
\frac{d\sigma(\lambda= +\frac{1}{2})}
{ dz_1 d\cos\theta_1 dz_2 d|{\bf p}_{2\perp}|^2 d\omega } 
+
\displaystyle
\frac{d\sigma(\lambda= -\frac{1}{2})}
{ dz_1 d\cos\theta_1 dz_2 d|{\bf p}_{2\perp}|^2 d\omega } 
}. \label{defi} 
\end{equation}
Substituting Eq. (\ref{cs}) into (\ref{defi}), one has 
\begin{equation} 
A_L=\frac{2|{\bf p}_{2\perp}| |{\bf p}_1| \sin\theta_1 \sin\omega  \hat W }
{ 2Q^2 W_1+
|{\bf p}_1|^2 \sin^2\theta_1 W_2
+ |{\bf p}_{2\perp}| |{\bf p}_1| \sin 2\theta_1 \cos\omega  W_3+ 
|{\bf p}_{2\perp}|^2(1- \sin^2\theta_1 \cos^2\omega  )W_4}.
\end{equation} 

 Our positivity analysis has useful  
phenomenological implications  to $A_L$.  
Let us assume $W_3=0$ in (\ref{l0}). Then,  we have 
\begin{equation} 
|{\bf p}_{2\perp}| |{\bf p}_1| |\hat W|\leq 
\sqrt{(Q^2W_1 +|{\bf p}_{2\perp} |^2 W_4)(Q^2 W_1+|{\bf p}_1 |^2W_2)}. 
\label{!}
\end{equation} 
Accordingly, there is  
\begin{equation} 
|A_L|\leq  \frac{2  
\sqrt{(Q^2W_1 +|{\bf p}_{2\perp} |^2 W_4)(Q^2 W_1+|{\bf p}_1 |^2W_2)} 
\sin\theta_1 \sin\omega  }
{ 2Q^2W_1+
|{\bf p}_1|^2 \sin^2\theta_1 W_2
+ |{\bf p}_{2\perp}| |{\bf p}_1| \sin 2\theta_1 \cos\omega  W_3+ 
|{\bf p}_{2\perp}|^2(\sin^2\theta_1 \cos^2\omega  -1)W_4}.
\label{bound}
\end{equation} 

 To further bound the discovery opportunity  in 
experimental  measurement, we now input into (\ref{bound}) 
the Callan-Gross  relation 
\begin{equation}
W_1+ \frac{z_1^2}{4}W_2=0 \label{callan}
\end{equation} 
for the process considered. 
[Making use of the quark-parton model, one can  
verify  Eq. (\ref{callan})  straightforwardly.] 
As a reasonable approximation, one can neglect all 
the $1/Q$-power suppressed effects, i.e, 
$W_3$- and $W_4$-terms in  the denominator of 
the right-hand side of (\ref{bound}).  As a  result,  we obtain 
the following upper bound to our single spin  asymmetry: 
\begin{equation} 
|A_L| \leq  \frac{4 M  \sin\theta_1 \sin\omega }
{z_1 Q (1+\cos^2\theta_1)}. \label{final} 
\end{equation}

It should be stressed  that (\ref{final}) is an amplified bound 
for $A_L$.  The reason is that in deriving (\ref{!}) 
from (\ref{l0}) and (\ref{sign}),  we have assumed $W_3=0$. 
Since both $W_3$ and $\hat W$ contribute at one-power 
suppressed level,  the  upper bound 
we derived is  quite  safe. Experimentally, the data 
on spin asymmetries are usually subject to  
large  statistical errors, so the above  positivity constraint
can be taken as a useful guide to judge the reliability
of data. On the other hand, (\ref{final})  also 
informs  us  that there is no large  
spin asymmetries  in some kinematic Al  regions. 
In addition,  our positivity constraints  can serve 
as  a consistency check for the future model 
calculations, if any.

To be illustrative, we consider the  two-pion production. 
 We  plot in Fig. 1 
 the  upper bounds of $|A_L|$ versus the 
beam-referenced production angle $\theta_1$  of the first pion. 
In  doing so,   we  take $Q=4.25$ GeV at which 
a longitudinally polarized electron-positron 
collider \cite{tau} is expected. Furthermore, 
we assume the ${\bf p}_1 \times {\bf p}_2$ plane is
perpendicular to the production plane of the first 
pion determined by ${\bf k}_1 \times {\bf p}_1$ so as 
to obtain the most optimistic upper bound. 
To guarantee  the Callan-Gross relation is reliable, 
we take the  energy  of the first pion to be  ten times its mass, 
which corresponds to $z_1\approx 0.66$.

 In summary,  the final-state interaction effects on 
the inclusive two-particle production is 
investigated in electron-positron annihilation within
the context of the one-photon annihilation 
approximation.  Such effects are characterized by  
$one$ structure function
in the Lorentz decomposition of the hadronic  tensor. 
Making use of the positivity of the hadronic  tensor, 
we derived an  upper bound for this structure function.
To access it experimentally, one 
needs to polarize the beam electron  longitudinally and 
measure the corresponding  single spin asymmetry. 
Further,  we presented an upper bound for the 
considered single spin asymmetry  by combining the 
Callan-Gross relation with the positivity constraint.

{\acknowledgements} One of the authors (W.L.) thanks 
J. Soffer for useful correspondence.

\newpage 

\centerline{\bf \huge Figure Caption}

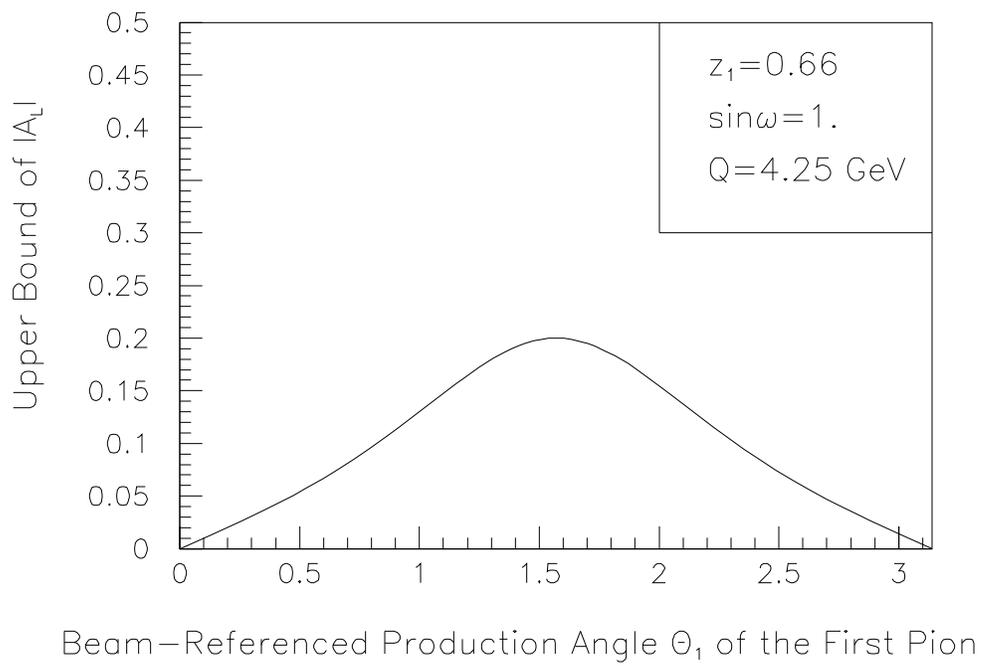
Figure 1.  Plot of the upper bound  of the  spin asymmetry 
($A_L$) for $ e^-(k_1,s_1) +e^+(k_2) \to \gamma^\ast \to \pi^+(p_1)
+ \pi^-(p_2)+X$ 
  versus the  beam-referenced outgoing angle ($\theta_1$) of the 
first pion.  To obtain the most optimistic upper bound, 
we assume the ${\bf p}_1 \times {\bf p}_2$ plane is
perpendicular to the production plane of the first 
pion determined by ${\bf k}_1 \times {\bf p}_1$. 
The  c.m. energy is set to be $Q=4.25$ GeV, 
at which a longitudinally polarized electron-positron 
collider  is  expected for the future Tau-Charm Factory \cite{tau}. 
To guarantee  the Callan-Gross relation is reliable, 
we take the  energy  of the first pion to be  ten times its mass, 
which corresponds to $z_1\approx 0.66$.

\end{document}